\documentclass{aipproc}
\layoutstyle{6x9}
\usepackage{pstricks,pst-plot}
\usepackage{amssymb}
\psset{unit=1cm,linewidth=0.7pt,arrowscale=1.5}

\newcommand\be{\begin{equation}}
\newcommand\ee{\end{equation}}
\def\one{\psscalebox{1.2}{\bf 1}}
\def\zero{\psscalebox{1.2}{\bf 0}}
\newcommand\aq{{\vphantom{Q}\smash{\tilde Q}}}
\newcommand\vev[1]{\left\langle #1\right\rangle}
\newcommand\tr{\mathop{\rm tr}\nolimits}

\begin{document}

\title[MSB in a Cooling Universe]{%
	Metastable Supersymmetry Breaking\break in a Cooling Universe%
	\footnote{Plenary talk at Pascos--07 conference
	   (Imperial College (London), July 2007),
		based on \cite{Texas}
		 (by W.~Fischler, myself,
       C.~Krishnan, L.~Manelli, and M.~Torres)
       and on unpublished work by M.~Torres and myself.
       }}
	
\author{Vadim S.\ Kaplunovsky}{%
	address={Physics Theory Group, University of Texas\\
		1 University Station, C1608, Austin, TX~78712, USA},
   email={vadim@physics.utexas.edu}}
   

\begin{abstract}
I put metastable supersymmetry breaking in a cosmological context.
I argue that under reasonable assumptions, the cooling down early Universe
favors metastable SUSY-breaking vacua over the
stable supersymmetric vacua.
To illustrate the general argument, I analyze the early--Universe
history of the Intriligator--Seiberg--Shih model.
\end{abstract}

\maketitle

Let me start with a historical note.
The metastable supersymmetry breaking (MSB) is actually quite old ---
Michael Dine and Willy Fischler \cite{DF} constructed interesting
models with both SUSY and non-SUSY vacua back in 1981.
But later, when people searched for SUSY breaking
driven by strong interactions (in a UV-free
but IR-strong hidden sector) but didn't have techniques for
analyzing effective potentials in strongly interacting theories,
they focused on models where SUSY {\it had to break} because
there were no SUSY vacua at all, and no runaway directions \cite{ADS}.
Although many new techniques for analyzing IR-strong gauge theories
emerged in mid-nineties, the search for SUSY breaking remained
focused on {\it true vacua} (lowest-energy states) without SUSY.
It took the Intriligator--Seiberg--Shih paper \cite{ISS} to bring
the MSB back into limelight.

Following Intriligator, Seiberg and Shih,
there was a flood \cite{Franco:2006es}--\cite{Cho:2007yn}\footnote{%
    As of this writing, the arXiv has over 130 papers on the subject,
    but I cannot quote them all because of space limitations.
    The papers \cite{Franco:2006es}--\cite{Cho:2007yn}
    are just a small sample of this flood.%
    }
of metastable SUSY--breaking (MSB) models,
many of them string based.
Such models have multiple vacua, some supersymmetric and some
SUSY-breaking; sometimes there also supersymmetric runaway
directions.
The physically-interesting non-SUSY vacua are metastable but
very long lived.
Given infinite time, they would eventually tunnel to a SUSY
vacuum or a runaway state, but this takes
much longer then the present age of the Universe.
So if a model somehow ended in the metastable state
soon after the Big Bang, it would stay there until today
and long afterwards.

Naturally, this raises {\bf the Big Question: \it Given an MSB
model with multiple vacua, which vacuum would be selected
by the cosmological history of the Early Universe?}
In this talk I am presenting {\bf Our Answer: \sl
Under reasonable assumptions, \it the Early Universe
favors the metastable SUSY--breaking vacua.}

Let me start by summarizing the common features of MSB models
which can be used as SUSY-breaking hidden sectors of phenomenologically
viable theories.
\begin{itemize}
\item
For phenomenological reasons, the scale of SUSY breaking should
be either $10^5$--$10^6$~GeV (for the direct gauge mediation of SUSY
breaking to the~Standard Model),
or $10^{10}$--$10^{11}$~GeV (for the indirect gauge mediation, or for
the SUGRA+K\"ahler mediation).
In any case, the SUSY breaking itself (as opposed to its mediation)
does not depend on SUGRA effects and can be approximated by the
rigid SUSY.

\item
The model must have an approximate $U(1)_{\bf R}$ symmetry
to facilitate the spontaneous SUSY breakdown.
I am not sure if this R-symmetry is quite as necessary as
Seiberg {\it et~al} claim \cite{Claim}, but it certainly helps,
and thus far all known
MSB models do have an approximate $R$--symmetry.

\item
In order to give masses to the Standard Model's gauginos,
the R-symmetry must be broken.
Usually, a small {\it explicit} breaking of the R-symmetry
is amplified via spontaneous breaking.
Alternatively, a small {\it explicit} breaking of the R-symmetry
in the SUSY-breaking hidden sector is amplified in the mediator sector.
But a purely spontaneous R-symmetry breaking would be bad because of
exactly-massless Goldstone bosons.

\item
Explicit breaking of the R-symmetry leads to additional vacua
with unbroken SUSY.
For {\it small} R-symmetry breaking, those SUSY vacua are far
away (in field space) from the non-SUSY vacua.
That is, the scale $\sigma$ of VEVs and masses in the SUSY vacuum
is much bigger then the scale $\mu$ in the non-SUSY vacuum,
\be
{\sigma\over\mu}\ \sim\ \Bigl(\hbox{R-symmetry breaking}\Bigr)
    ^{\rm some\, negative\, power}\
\gg\ 1.
\ee

\item
The non-SUSY vacuum is metastable because it
has higher energy density then the SUSY vacua.
{\it But for $\sigma\gg\mu$, its lifetime is very long.}
Indeed, the potential barrier between the SUSY and non-SUSY
vacua is very wide, $\Delta\Phi=O(\sigma)$,
while the potential difference is only $\Delta V=O(\mu^4)$.
The tunneling action of a Euclidean bubble of the true vacuum inside
the false vacuum is
\be
S\ \sim\ {V^2\,(\Delta\Phi)^4\over (\Delta V)^3}\
\gtrsim\ {(\Delta\Phi)^4\over\Delta V}\ \sim\
\left({\sigma\over\mu}\right)^4\ \gg\ 1.
\ee
Thus, {\sl for $\sigma\gtrsim 10\,\mu$, the metastable SUSY-breaking
vacuum would easily survive until the present age of the Universe}.

\end{itemize}

\noindent
To place an MSB model in a broader context, I make the following
{\bf assumptions}:
\begingroup
\renewcommand\labelitemi{$\star$}
\begin{itemize}
\item
The SUSY-breaking hidden sector has nothing to do with
inflating the Universe.
The Inflation happens due to dynamics of a completely separate
sector of the overall theory.

\item
The overall theory has yet another sector, which cancels the
cosmological constant due to SUSY breaking in the metastable vacuum.

\item
After the Inflation, the reheating temperature is high enough
for the high-temperature phase of the SUSY-breaking sector,
\be
T_{\rm reheat}\ >\ O(\sigma)\ \gg\ \mu.
\label{Reheat}
\ee

\end{itemize}
\endgroup
\noindent
I claim that under these assumptions, the cosmological evolution
of the MSB sector during the early Universe tends to end up
in the metastable non-SUSY vacuum state.
Here is the {\bf basic argument}:

After the Inflation is over, the Hubble expansion of the Universe
is slow ($H\ll T$) and the temperature decreases slowly enough
for the quasistatic approximation:
At any given time, the fields and particles are
in thermal equilibrium for the
appropriate temperature, and the free energy is minimized.
Or rather, the free energy density $\cal F$
is always in a {\it local} minimum.

Multiple local minima of $\cal F$ correspond to multiple phases:
one stable (the global minimum), the others metastable.
Transitions between the phases require tunneling or thermal
activation, and can be very slow.
If they take longer then the Hubble time, they never happen,
and the SUSY-breaking sector stays in a metastable phase.

The non-SUSY phase has higher potential energy then the ``SUSY''
phase\footnote{%
   By ``SUSY'' phase I mean the phase which for zero temperature
   reduces to the SUSY vacuum.
   At finite temperatures SUSY is broken, hence the quote marks.%
   }
but also higher entropy (because it has lighter particles,
$\mu\ll\sigma$).
At higher temperatures, the entropy wins over the potential energy,
which favors the non-SUSY phase.
And at very high temperatures ($T>O(\sigma)$) the SUSY phase
disappears altogether, because the slope of the entropy function
overwhelms the minimum of the scalar potential.

I assume the Universe reheats to $T\gg\sigma$ and then slowly
cools down.
At first, the MSB sector has only the non-SUSY phase.
As the temperature drops below $O(\sigma)$, other phases develop,
but the non-SUSY phase has lowest free energy,
and the sector remains in that phase.

Much later, for $T=O(\mu)$, the scalar potential wins over entropy,
and the non-SUSY phase becomes metastable, while the SUSY phase
becomes thermodynamically stable.
But the first-order transition from the metastable to the stable
phase requires either tunneling or thermal activation of a bubble,
and both processes are very slow for $\sigma\gg\mu$:
\begin{eqnarray}
\Gamma_{\rm tunneling} &\sim&
\exp\left(-S[\mbox{Eucl. 4D bubble}]\right)\ \ll\ 
\exp\left( -O(1)\times\left({\sigma\over\mu}\right)^4\right),
\label{G4}\\
\Gamma^{\rm thermal}_{\rm activation} &\sim&
\exp\left(-{E[\mbox{3D bubble}]\over T}\right)\ \ll\ 
\exp\left( -O(1)\times\left({\sigma\over\mu}\right)^3\right).
\label{G3}
\end{eqnarray}
Thus today, 13.5 gigayears since the temperature crossed the
transition point, the theory remains in the metastable SUSY-breaking phase,
and will stay there for many more gigayears.

To illustrate this general argument with a specific example,
let us consider the {\bf Intriligator--Seiberg--Shih model} \cite{ISS}.
In that model, the {\bf UV} theory is simply SQCD with massive but light
quarks, $m_q\ll\Lambda$, and $N_c<N_f<{3\over2}N_c$.
The {\bf IR} theory at energies below $\Lambda$ follows by
Seiberg duality: it's SQCD with $N=N_f-N_c$ colors,
$N_f>3N$ flavors, and $N_f^2$ extra gauge singlets $\Phi_{ff'}$.
The singlets originate from the mesons of the UV theory,
$\Phi_{ff'}=\Lambda^{-1}\vev{\tilde q_f q_{f'}}$;
their flavor quantum numbers are $\bf Adj+1$.
The IR quarks $Q_f$ and antiquarks $\aq_f$
and the IR gauge fields
do not have clear UV origins.
The superpotential is
\be
W\ =\ h\tr(\Phi \aq Q)\ -\ h\mu^2\tr(\Phi)\
+\ hNC\Bigl( \det(\Phi)\Bigr)^{1/N}
\label{WISS}
\ee
where $\mu^2\simeq \Lambda m_q\ll\Lambda^2$, and
$C\simeq \Lambda^{3-(N_f/N)}$.
The K\"ahler function is approximately canonical
(modulo perturbative renormalization),
because the theory is IR free, ${\beta_h,\beta_g>0}$.

Without the non-renormalizable third term, there is
exact $U(1)_{\bf R}$ symmetry, and SUSY has to break:
\be
F_\Phi\, \propto\, \aq Q\, -\, \mu^2\times\one_{N_f\times N_f}\,
\neq\, 0\quad \hbox{because}\
\mathop{\rm rank}(\aq Q)\le N<N_f\,.
\ee
In the non-SUSY vacuum,
\be
\vev\Phi\,=\,0,\qquad\vev{Q}\,=\,\vev{\aq}^\top\,=\,\mu\times
\left(\vcenter{\ialign{$#$\hfil\cr
	\one_{N\times N}\cr
	\zero_{N\times(N_f-N)}\cr
    }}\right) .
\ee

The determinant term in $W$ breaks the R-symmetry, and leads
to an additional SUSY vacuum (or rather $N_f-N=N_c$ vacua) with
\begin{eqnarray}
\vev{Q}\,=\,\vev{\aq}\,=\,0,&&\qquad \vev\Phi\,=\,\sigma\times
\one_{N_f\times N_f}\,,\\
\sigma\,=\,(\mu^2/C)^{N/(N_f-N)}&\simeq&
\left(\mu^{2N}\Lambda^{N_f-3N}\right)^{1/(N_f-N)}\,\gg\,\mu .
\end{eqnarray}

To analyze and depict various phase transitions in this model, 
I am restricting its fieldscape to a two-parameter ansatz:
\be
\Phi\,=\,\varphi\times\one_{N_f\times N_f}\,,\qquad
Q\,=\,\aq^\top\,=\,q\times
\left(\vcenter{\ialign{$#$\hfil\cr
	\one_{N\times N}\cr
	\zero_{N\times(N_f-N)}\cr
    }}\right) ,
\ee
and real $\varphi$ and $q$ (for real $\sigma$ and $\mu$).
In this ansatz, the tunneling from the non-SUSY to the SUSY vacuum
happens along the following path:
$$
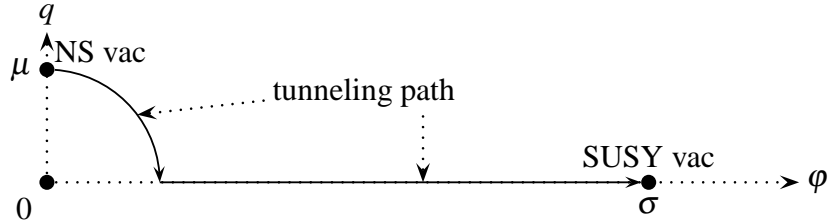
\begin{pspicture}(-0.5,-0.5)(10.5,2.5)
\psline[linestyle=dotted,linewidth=1pt]{->}(0,0)(10,0)
\rput[l](10.1,0){$\varphi$}
\psline[linestyle=dotted,linewidth=1pt]{->}(0,0)(0,2)
\rput[b](0,2.1){$q$}
\pscircle*(0,1.5){0.1}
\rput[r](-0.2,1.5){$\mu$}
\rput[bl](0.1,1.6){NS vac}
\pscircle*(8,0){0.1}
\rput[t](8,-0.2){$\sigma$}
\rput[b](8,0.2){SUSY vac}
\pscircle*(0,0){0.1}
\rput[rt](-0.2,-0.2){0}
\psarc{<-}(0,0){1.5}{0}{86}
\psline{->}(1.5,0)(7.9,0)
\rput[bl](3,1){tunneling path}
\psline[linestyle=dotted,linewidth=1pt]{->}(3,1.1)(1.2,0.9)
\psline[linestyle=dotted,linewidth=1pt]{->}(5,1)(5,0)
\end{pspicture}
$$

At zero temperature, the effective potential of the model
is approximately
\be
V(\varphi,q)\ \approx\ 
{N_fh^2\mu^4\over Z_\Phi(\varphi)}\times
    \left(1-(\varphi/\sigma)^{(N_f/N)-1}\right)^2\
+\ Nh^2\left(q^4-2\mu^2 q^2+2\varphi^2 q^2\right)
\ee
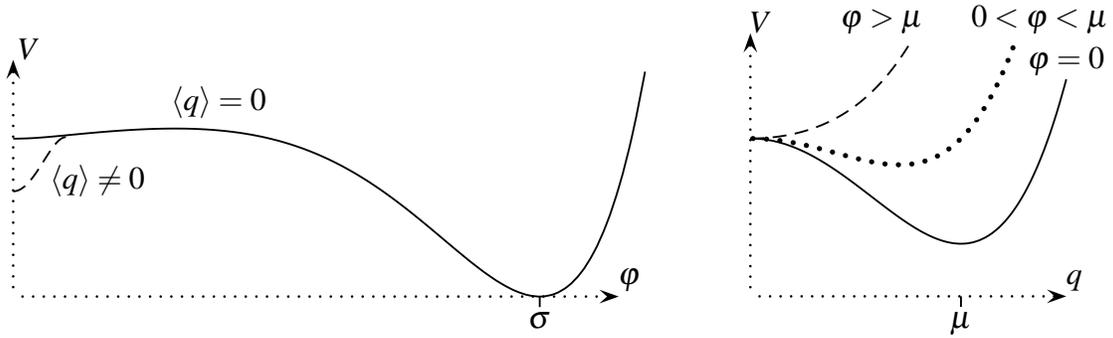
$$
\psset{unit=7mm}
\begin{pspicture}(0,-1.5)(21,5)
\rput(14,0){%
	\psline[linestyle=dotted,linewidth=1pt]{->}(0,-1)(6,-1)
	\rput[bl](6,-0.9){$q$}
	\psline(4,-1)(4,-1.2)
	\rput[t](4,-1.3){$\mu$}
	\psline[linestyle=dotted,linewidth=1pt]{->}(0,-1)(0,4)
	\rput[bl](0,4){$V$}
	\psplot[plotstyle=curve]{0}{6}%
	  {x dup mul 32 sub x dup mul mul 128 div 2 add}
	\rput[b](6,3.2){$\varphi=0$}
	\psplot[plotstyle=curve,linestyle=dotted,linewidth=2pt]{0}{5}%
	  {x dup mul 16 sub x dup mul mul 128 div 2 add}
	\rput[b](5.5,4){$0<\varphi<\mu$}
	\psplot[plotstyle=curve,linestyle=dashed]{0}{3}%
  {x dup mul 16 add x dup mul mul 128 div 2 add}
	\rput[b](2.5,4){$\varphi>\mu$}
	}
\rput(0,-1){%
	\psline[linestyle=dotted,linewidth=1pt]{->}(0,0)(11.5,0)
	\rput[lb](11.5,0.1){$\varphi$}
	\psline[linestyle=dotted,linewidth=1pt]{->}(0,0)(0,4.5)
	\rput[lb](0.1,4.5){$V$}
	\psline(10,0)(10,-0.2)
	\rput[t](10,-0.3){$\sigma$}
	\psplot[plotstyle=curve]{0}{12}%
	  {x dup mul 1 add log 4 div 1 exch sub -.272727 exp %
	    x 10 div dup mul dup mul 1 sub dup mul mul 3 mul}
	\rput[bl](3,3.4){$\vev{q}=0$}
	\psplot[plotstyle=curve,linestyle=dashed]{0}{1}%
	  {x  dup mul 1 sub dup mul -1.03 mul 3.03 add}
	\rput[tl](0.7,2.5){$\vev{q}\neq0$}
	}
\end{pspicture}
$$
At finite temperatures, the effective potential --- $i.\,e.$,
the free energy density --- comprises
\begin{eqnarray}
{\cal F}(\varphi,q) &=&
V(\varphi,q)\ +\ {\cal F}^{\rm 1\,loop}_T(\varphi,q)\
  +\ \mbox{higher order corrections},\\
{\cal F}^{\rm 1\,loop}_T &=&
T\times\!\int\!{d^3p\over8\pi^3}\,\mathop{\rm Str}
	\left( 1\,-\,(-1)^F\exp\left(-\sqrt{p^2+M^2}/T\right)\right),\\
&& \mbox{where the spectrum of $M^2$
  depends on $\varphi$ and $q$.}\nonumber
\end{eqnarray}
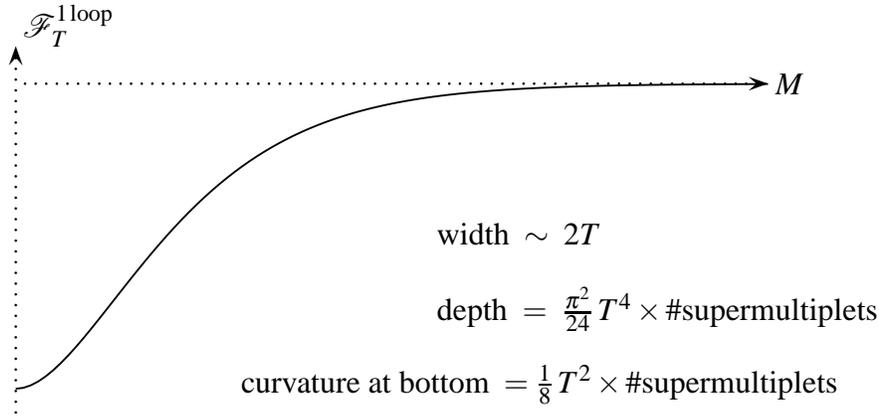
$$
\begin{pspicture}(0,-4.5)(12,+1)
\psline[linestyle=dotted,linewidth=1pt]{->}(0,0)(10,0)
\rput[l](10.1,0){$M$}
\psline[linestyle=dotted,linewidth=1pt]{->}(0,-4.5)(0,+0.5)
\rput[lb](0.1,+0.5){${\cal F}^{\rm 1\,loop}_T$}
\fileplot[plotstyle=line]{fe.points}
\rput[l](5.6,-2){width ${}\sim\ 2T$}
\rput[l](5.6,-3){depth ${}=\ {\pi^2\over24}\,T^4\times\#{}$supermultiplets}
\rput[l](3,-4){curvature at bottom ${}={1\over8}\,T^2\times\#{}$supermultiplets}
\end{pspicture}
$$

For high temperatures, $T\gg\sigma$,
the thermal energy ${\cal F}^{1\,\rm loop}_T$
completely overwhelms the scalar potential $V$.
Consequently, the net free energy ${\cal F}(\varphi,q)$ has only one
minimum at $\varphi=q=0$, which means there is a unique
high-temperature phase HT.
Note that this phase  is  distinct from the non-SUSY
phase at low temperatures because of different squark expectation values
($\vev{q}=0$ in the HT phase versus $\vev{q}=\mu$ in the
low-temperature NS phase).

For medium temperatures, $\mu\ll T\ll\sigma$,
the thermal energy overwhelms the scalar potential for
$q,\varphi\lesssim\mu\ll T$.
But for $h\varphi\gg T$, the ${\cal F}^{1\,\rm loop}_T(\varphi)$
flattens out
(because all masses are either much larger or much smaller
than $T$), so the minimum of $V$ at $\varphi=\sigma$ becomes
visible in the overall free energy:
\par
$$
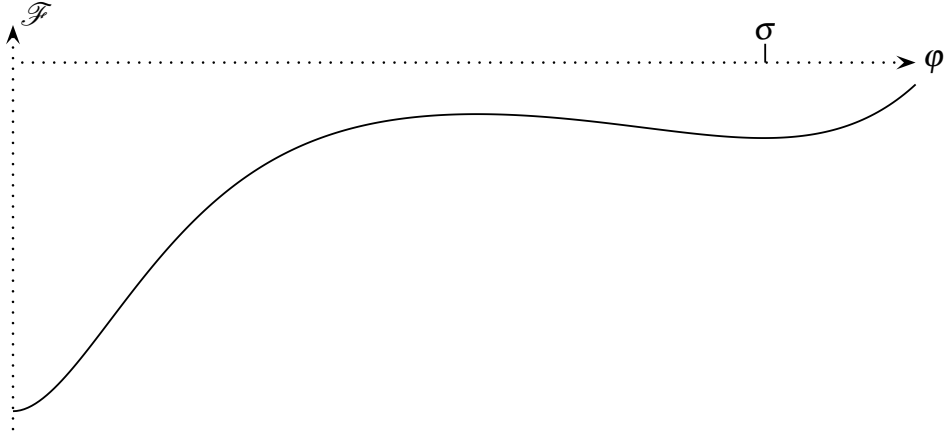
\psset{xunit=10mm,yunit=5mm}
\begin{pspicture}(0,-8)(12.5,+4)
\psline[linestyle=dotted,linewidth=1pt]{->}(0,2)(12,2)
\rput[l](12.1,2){$\varphi$}
\psline(10,2)(10,2.5)
\rput[b](10,2.6){$\sigma$}
\psline[linestyle=dotted,linewidth=1pt]{->}(0,-8)(0,+3)
\rput[lb](0.1,+3){${\cal F}$}
\fileplot[plotstyle=line]{f2.points}
\end{pspicture}
$$
This gives us two phases: the stable HT phase with $\vev{q}=\vev\varphi=0$,
and the metastable ``SUSY'' phase
with $\vev{q}=0$ and $\vev\varphi\approx\sigma$.
{\it In the cooling Universe, the system is in the HT phase
before temperature drops below $O(\sigma)$, and afterwards
it remains in the HT phase because it's stable.}

As the Universe cools down further, the energy difference between
the HT and the ``SUSY'' phases becomes smaller, and eventually
changes sign at the critical temperature
\be
T_c^\phi\ \approx\ {2\mu\over\sqrt{N}}\times
\root4\of{6\,{Nh^2\over 8\pi^2}}
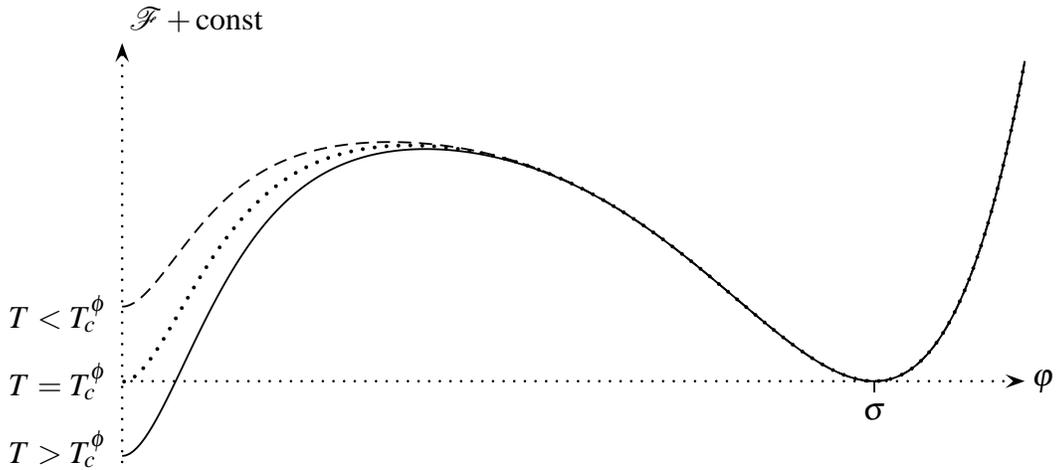
\ee
$$
\psset{xunit=10mm,yunit=30mm}
\begin{pspicture}(-2,-0.4)(12.5,+1.7)
\psline[linestyle=dotted,linewidth=1pt]{->}(0,0)(12,0)
\rput[l](12.1,0){$\varphi$}
\psline(10,0)(10,-0.05)
\rput[t](10,-0.1){$\sigma$}
\psline[linestyle=dotted,linewidth=1pt]{->}(0,-0.4)(0,+1.5)
\rput[lb](0.1,+1.55){${\cal F}+\rm const$}
\fileplot[linestyle=solid,plotstyle=line]{f3.points}
\rput[r](-0.2,-0.3){$T>T_c^\phi$}
\fileplot[linestyle=dashed,plotstyle=line]{f5.points}
\rput[r](-0.2,0){$T=T_c^\phi$}
\fileplot[linestyle=dotted,linewidth=1.5pt,plotstyle=line]{f4.points}
\rput[r](-0.2,+0.3){$T<T_c^\phi$}
\end{pspicture}
$$
For temperatures below $T_c^\phi$, the HT phase becomes
metastable while the ``SUSY'' phase becomes thermodynamically stable.
{\it Nevertheless, the model remain in the now-metastable HT phase
because the first order transition between the two phases
is extremely slow.}

At somewhat lower temperature
\be
T_c^q\ \approx\ {2\mu\over\sqrt{N_f+2N}}\
\sim\ (0.4\ {\rm to}\ 0.75)\times T_c^\phi
\ee
there is another phase transition in the squark direction:
$$
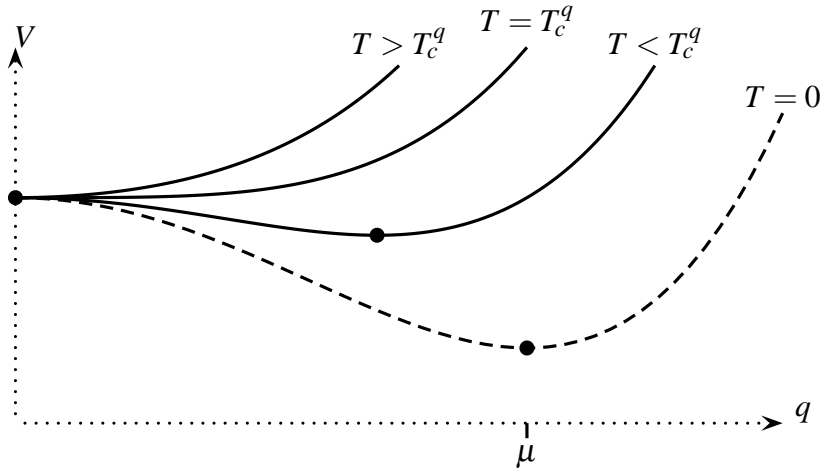
\psset{xunit=17mm,yunit=10mm,runit=10mm,linewidth=1.2pt}
\begin{pspicture}(0,-1.5)(6.5,4.5)
\psline[linestyle=dotted]{->}(0,-1)(6,-1)
\rput[bl](6.1,-1){$q$}
\psline(4,-1)(4,-1.2)
\rput[t](4,-1.3){$\mu$}
\psline[linestyle=dotted]{->}(0,-1)(0,4)
\rput[bl](0,4){$V$}
\psplot[plotstyle=line,linestyle=dashed]{0}{6}%
  {x dup mul 32 sub x dup mul mul 128 div 2 add}
\rput[b](6,3.2){$T=0$}
\pscircle*(4,0){0.1}
\psplot[plotstyle=curve]{0}{5}%
  {x dup mul 16 sub x dup mul mul 128 div 2 add}
\rput[b](5,3.85){$T<T_c^q$}
\pscircle*(2.828,1.5){0.1}
\psplot[plotstyle=curve]{0}{4}%
  {x dup mul  dup mul 128 div 2 add}
\rput[b](4,4.15){$T=T_c^q$}
\psplot[plotstyle=curve]{0}{3}%
  {x dup mul 16 add x dup mul mul 128 div 2 add}
\rput[b](3,3.85){$T>T_c^q$}
\pscircle*(0,2){0.1}
\end{pspicture}
$$
This transition is second order, and proceeds without delay.
As soon as the Universe cools down to $T_c^q$,
the HT phase with $\vev{q}=\vev\varphi=0$ disappears,
and the model enters the low-temperature non-SUSY phase NS with
$\vev\varphi=0$ but $\vev{q}\neq0$.

Similar to the HT phase below $T_c^\phi$,
{\it the NS phase is metastable.}
Given infinite time, it would eventually decay into the ``SUSY'' phase
with $\vev{q}=0$ and $\vev\varphi\approx\sigma$.
But for $\sigma\gtrsim 20\mu$,
the tunneling and the thermal activation are both very slow
--- {\it cf.}\ eqs.~(\ref{G4}--\ref{G3}) ---
and the decay takes longer then the present age of the Universe.

Instead of decay, the model remains in the metastable NS phase.
As the temperature drops, the squark VEV grows toward $\vev{q}=\mu$,
and {\it the model cools down to the non-SUSY vacuum}.

Besides the Intriligator--Seiberg--Shih model, M.~Torres and I
have analyzed similar models with weakly gauged flavor symmetries
(the whole $SU(N_f)_V$ or its subgroups).
Such models have spontaneously broken R-symmetry at $T=0$ and more
complicated phase structures at $t>0$.
But of the end of the evolution, they too end up in metastable non-SUSY
vacuum states.

\underline{To summarize our results}, {\bf Metastable SUSY breaking is OK}.
In models with both non-SUSY and SUSY vacua where the latter have
much larger VEVs and masses then the former,
this little hierarchy not only keeps the metastable SUSY-breaking
vacua very long lived, but also leads the cosmological evolution of the model
toward those vacua.
{\bf But the devil is in details}:
\begin{itemize}
\item
Above all, the model must work!
And mediation of SUSY breaking to the SSM should also work.
\item
There should be no way {\it around} the potential barrier between the
vacua. The pseudo-moduli directions are particularly dangerous.
\item
The phase diagram of the model should direct its thermal evolution toward
the desired non-SUSY vacuum.
In models with several distinct vacua, this could be quite a challenge.
\item
The mediators should not screw things up.
\item
{\it Etc., etc.} \dots

\end{itemize}

\medskip
\noindent {\bf Acknowledgements:}
Simultaneously with our paper \cite{Texas} on which this talk is based,
two other groups published independent papers \cite{California,England}
with similar conclusions.
\newline
The research of our group was supported by the NSF under grant PHY--0455649.


\end{document}